# Nanoscale Bandgap Tuning across Inhomogeneous Ferroelectric Interface

*Jing Wang[1,#], Houbing Huang[2,#], Wangqiang He[2], Qinghua Zhang[3], Danni Yang[1], Yuelin Zhang[1], Renrong Liang[5], Chuanshou Wang[1], Xingqiao Ma[2], Lin Gu[4], Longqing Chen[6], Ce-Wen Nan[3] and Jinxing Zhang[1,*]*

1, Department of Physics, Beijing Normal University, 100875, Beijing, China

2, Department of Physics, University of Science and Technology Beijing, 100083, Beijing, China

3, School of Materials Science and Engineering, Tsinghua University, 100084, Beijing, China

4, Institute of Physics, Chinese Academy of Science, 100190, Beijing, China

5, Tsinghua National Laboratory for Information Science and Technology, Institute of Microelectronics, Tsinghua University, 100084, Beijing, China

6, Department of Materials Science and Engineering, The Pennsylvania State University, University Park, PA 16802, USA

ABSTRACT: We report the nanoscale bandgap engineering via a local strain across the inhomogeneous ferroelectric interface, which is controlled by the visible light-excited probe voltage. Switchable photovoltaic effects and spectral response of the photocurrent were explored to illustrate the reversible bandgap variation (~0.3 eV). This local-strain-engineered bandgap has been further revealed by *in-situ* probe-voltage-assisted valence electron energy loss spectroscopy (EELS). Phase-field simulations and first-principle calculations were also employed to illustrate the large local strain and the bandgap variation in ferroelectric perovskite oxides. This reversible bandgap tuning in complex oxides demonstrates a framework to understand the optical-related behaviors (photovoltaic, photo-emission, and photo-catalyst effects) affected by order parameters such as charge, orbital and lattice.

# INTRODUCTION

Bandgap engineering of semiconductors has attracted people's interests for the past decades due to the emerging quantum phenomena [1-2] and the applications in photocatalysis [3], solar cells [4] and other optoelectronic devices (e.g., photoemitters [5], photodetectors [6] and photomodulators [7]). The bandgap of semiconductors can be successfully engineered by employing quantum confinement effects [1-2], chemical doping [8], epitaxial strain [9-10], bending effects (free-standing carbon nanotube [11] and $MoS_2$ monolayer [12] etc.) or hetero-structures [13-14]. However, a reversible control of the bandgap under the external stimulus in integrated semiconductors may be challenging. Therefore, the exploration of bandgap engineering in new material systems is highly desired. In strong correlated materials, the optical band structures are coupled with other order parameters, where the bandgap could be modulated under external stimuli (stress, optical excitation and electric/magnetic fields etc.).

Multiferroic perovskite oxides exhibit the coexistence of ferro/piezoelectricity, magnetism and ferroelasticity due to the strong correlation between lattice and other degrees of freedom (e.g., charge, orbital and spin), which provide people potential candidates to study the electronic and optical properties under the application of external stimuli. Dong *et al*. and Wang *et al*. examined bandgap change in ferroelectric $BiFeO_3$ (BFO) [15] and $KNbO_3$ [16] via a giant external compressive strain. Although such a large strain (over several percent) cannot be realized in bulk crystals in practice, these studies theoretically demonstrate that the electronic band structures in ferroelectric materials could be reversibly modulated by a compressive mechanical stimulus. Therefore, it gives us a strong push to find a way to reversibly input a large external controllable strain (compressive) in ferroelectric oxides without breaking the crystals.

The distinct structural deformation and the consequent large local strain in multiferroic materials can be achieved by the application of a mechanical force [17] or electric field [18-19] on the nanoscale probe. This probe-assisted large local strain can be reversibly achieved without breaking the crystal due to the gradual release of internal stress. In this work, we explored the reversible nanoscale bandgap tuning via a local strain across the inhomogeneous ferroelectric interface, which is controlled by the visible light-excited probe voltage. Switchable photovoltaic effects and spectral response of the photocurrent were explored by photoexcitation-assisted atomic force microscopy (P-AFM) technique to illustrate the reversible bandgap variation (~0.3 eV). This local-strain-engineered bandgap has been further revealed by *in-situ* probe-voltage-assisted valence electron energy loss spectroscopy (EELS). Phase-field simulations and first-principle calculations were also employed to illustrate the large local strain and the bandgap variation in ferroelectric perovskite oxides.

## RESULTS AND DISCUSSION

**A good ferroelectric model system.** With a high ferroelectric and antiferromagnetic transition temperature (i.e., $T_C = 830$ ℃ and $T_N = 370$ ℃) [20], BFO thin films with a rhombohedral (R)-like phase grown on 001-oriented SrTiO$_3$ (STO) substrates were used as a model system in this study. The intrinsic optical bandgap of BFO is ~2.7 eV [21], which exhibits efficient photovoltaic effect within visible light region [22-29]. In order to carry out the photo-electronic measurement, an epitaxial SrRuO$_3$ (SRO) was inserted as bottom electrodes. The topography, crystal structure, thickness and ferroelectric properties of the thin films were characterized with contact-mode AFM, X-ray diffraction (XRD), cross-sectional scanning electron microscopy (SEM) and piezoresponse force microscopy (PFM), which can be seen in supplementary Fig. S1. The detailed growth conditions, XRD, SEM characterizations, scanning-probe measurements and poling process of ferroelectric thin films can be seen in Materials and Methods.

**A probe voltage induced by visible light-excitation at the probe/film interface.** Local ferroelectric switching behaviors with the presence of visible light at the probe/film interface were studied to explore the interaction between the visible light and the local ferroelectric polarizations, which can be seen in Fig. 1(a). Fig. 1(b) shows the schematic of experimental setup for the local piezoresponse measurement. Firstly, a 180° out-of-plane polarization switching was acquired in BFO thin film to obtain the local switching bias (-1V and +2.8V respectively) without the illumination (black curve). When a visible light was applied at the probe/film interface, the switching bias reduced to -0.4V and +1.6V (red curve) respectively and then totally recovered (brown curve) when the light was removed. This observation indicates that the visible light can induce a probe voltage of 1.2V (yellow arrow) at the probe/film interfaces, which is opposite to the out-of-plane ferroelectric polarizations (green arrow) as shown in the magnified part of Fig. 1(b). Consequently, this probe voltage will give rise to a local strain in BFO thin film underneath the probe as shown in the magnified part of Fig. 1(b). The key question would be if this local strain will tune the band structure of the ferroelectric thin film.

**Switchable photovoltaic effects and local bandgap variation in BFO thin film based on a point-contact geometry.** The most straightforward way to detect the bandgap variation in ferroelectric thin films under this local strain is the local photo-electronic measurement. Fig. 2(a) and 2(b) shows the local ferroelectric photovoltaic effects in BFO thin film with out-of-plane downward and upward polarizations, respectively. The detailed measurements can be seen in Materials and Methods as well as supplementary Fig. S2. The open-circuit photovoltage ($V_{oc}$) and short-circuit photocurrent ($I_{sc}$) can be totally switched off when the light illumination was removed (yellow curves). The inserted out-of-plane PFM phase images of Fig. 2(a) and 2(b)) demonstrate the polarizations did not change after the local transport measurements. The strong photovoltaic effects have been repeatedly observed at multiple points and the results can be repeated for more than ten times at each point. This observation further indicates part of the light-excited charge carriers are captured by the surface defects, which enhances the surface charge screening and further stabilizes the ferroelectric polarizations as examined in the previous report [30]. While, a majority of the carriers are transported to the external circuit, such that a considerable $I_{sc}$ can be detected as shown in Fig. 2(a) and 2(b). On-off photocurrent as a function of time was also recorded to further confirm the effect of visible light excitation as shown in supplementary Fig. S3. The abrupt ascent (marked by red arrow) and descent (marked by blue arrow) of the photocurrent at the onset of light-on and -off region indicate the photovoltaic effects arise from the visible light excitation but not the thermal effect.

Based on the above characterization of local ferroelectric photovoltaic effects, spectral response of $I_{sc}$ [25] was recorded to explore the local bandgap in the point contact geometry. To make a parallel comparison, $I_{sc}$ as a function of wavelength of the illuminating light was investigated in the point-contact geometry (contact radius, ~10 nm [28]) and a capacitor structure (contact radius, ~5μm) respectively. As shown in the black curve of Fig. 2 (c), the highest $I_{sc}$ was detected at ~450 nm in the case of the capacitor structure, closely corresponding to the measured BFO bandgap of 2.75 eV [21][31]. While, in the point contact geometry (red curve), the highest $I_{sc}$ occurred at ~520 nm, which corresponds to a bandgap of 2.38 eV. These results indicate there is a bandgap drop of ~0.3 eV in BFO thin film underneath the probe compared with the case of capacitor structure. The detailed measurements for Fig. 2(c) can be seen in Materials and Methods and supplementary Fig.

S4. To fully understand the spectral responses of $I_{sc}$ in BFO thin films in the capacitor structure and the point-contact geometry, two schematics were drawn as shown in Fig. 2(d) and 2(e) respectively. Compared with the capacitor structure (Fig. 2(d)), there is a large local strain in BFO thin film with out-of-plane downward polarizations in the point-contact geometry (Fig. 2(e)), which is induced by the light-excited probe voltage with opposite direction compared with the polarizations. This probe-voltage induced local strain can be expected to tune the local band structure.

To further confirm this observation, *in-situ* probe-voltage-assisted valence EELSs were carried out in the point-contact geometry (Fig. 3). High resolution TEM image (Fig. 3(a)) indicates the high quality of the epitaxial thin films. Fig. 3(b) illustrates the *in-situ* experimental set-up, where a point-contact geometry was designed. The bandgap was derived by extrapolating the energy loss intensity from the onset region using a linear function [32]. Compared to the case of zero probe bias, the bandgap was decreased by 0.36 eV when a voltage of 1V (upward) was applied on the nanoscale probe as shown in the magnified part of Fig. 3(c), where the out-of-plane components of the polarizations were downward. This result further demonstrates the probe voltage can induce the local strain [19], which assists the tuning of local band structure in the ferroelectric thin films with point-contact geometry.

**Phase-field simulations and first-principle calculations.** In order to have a quantitative analysis of the probe-voltage induced local strain in BFO thin film underneath the probe, phase-field simulations were carried out with the application of probe voltage. A single domain with polarization along $[\bar{1}1\bar{1}]$ direction was simulated as shown in Fig. S5(a). Fig. S5(b) describes the polarization projection along c axis of BFO thin film with the application of 1.2V (upward) probe voltage in an out-of-plane downward domain state. The corresponding strain in BFO thin film was obtained as shown in Fig. 4(a). The total strain tensor trace ($\Sigma |s_{ii} - s_{ii}^{'}|$) was calculated to estimate the inhomogeneous strain variation, where $\Sigma s_{ii}$ and $\Sigma s_{ii}^{'}$ are strain tensor traces with and without the probe bias, respectively. A maximum out-of-plane compressive strain (~ -5.3%) underneath the probe was obtained, which is confined around the contact point and then decays significantly away from it. Similar results have been obtained in upward domain state when a probe voltage of 0.6V (downward) was applied as shown in Fig. S6. Fig. 4(b) and 4(c) shows the strain in BFO thin film as a function of the contact radiuses of 30 nm and 50 nm respectively. The maximum compressive strains decrease to -1.0% and -0.4%, respectively. The strain underneath the probe along the thickness direction as a function of contact radius was also simulated as shown in Fig. 5(d), indicating that the decreased inhomogeneity reduce the maximum local strain.

To further clarify this strain effect on the bandgap of R phase-like BFO, density functional theory (DFT) calculations were carried out as shown in Fig. 4(e). The calculations predicted the bandgap was 2.38 eV [33] at the ground state and decreased with the increasing compressive strain along [001] direction (green curve in Fig. 4(e)). However, since our GGA + U calculation underestimated the bandgap by 0.33 eV compared with experimental value [21], we shift the curve upward by this amount, as seen in the red curve of Fig. 4(e). The bandgap of BFO decreased ~0.12 eV when a -5.3% strain was applied along the [001] direction, which is consistent with our experimental observation of the bandgap drop from the spectral response of $I_{sc}$ (Fig. 2(c)) and *in-situ* EELSs (Fig. 3(c)) in BFO thin films with point-contact geometry, where a probe voltage (~1V, upward) was applied to induce the large local strain (~ -5.3%) as can be seen in the phase-field simulations (Fig. 4(a)). The detailed calculations and the evolution of the band structure of BFO thin film under different strains can be seen in Materials and Methods and Fig. S7.

In addition, the effect of contact force between the nanoscale probe and film surface has been taken into consideration. The magnitude of the contact force (75 nN) was extracted from the force curve as shown in Fig. S8, which would induce ~0.25 GPa stress ($\sigma$) in BFO thin film at the contact point considering the contact area (300 nm$^2$ [28]). The detailed measurement of the contact force can be seen in Materials and Methods. Taking into account the Young's modulus ($E$, ~169 GPa [34]) of R-phase BFO, the local strain ($\varepsilon$) generated by the contact force can be calculated as $\varepsilon = \sigma/E = 0.25$ GPa/169 GPa $\approx 0.1\%$, which is much less than the local strain (~5.3%) generated by the visible light-excited probe voltage. Therefore, the contact force between the nanoscale probe and the film surface will take negligible effect on the local strain and the consequent bandgap variation.

## CONCLUSION

To summarize, we demonstrated a reversible nanoscale bandgap engineering via a large local strain across the inhomogeneous ferroelectric interface, which is controlled by the visible light-excited probe voltage. This phenomenon has also been examined experimentally by *in-situ* probe-voltage-assisted EELSs. Phase-field simulations and first-principle calculations further illustrate the large local strain and the consequent bandgap drop in the ferroelectric perovskite oxides. This nanoscale bandgap tuning may help to better reversibly control the optical-related behaviors in complex oxides and provide the potential applications in next-generation optoelectronic devices.

## Materials and Methods

**Materials.** BFO thin films were grown on (001)-oriented STO substrates with SRO bottom electrodes using pulsed laser deposition (PLD). The growth temperature was maintained at 700 °C, and the working oxygen pressure was 15 Pa for both BFO and SRO thin films. For all films, a growth rate of 2.5 nm/min and a cooling rate of 5 °C/min under an oxygen atmosphere were used. A laser energy density of 1.2 J/cm$^2$ and a repetition rate of 3 Hz were used during the deposition. The thickness was controlled as ~30 nm and ~120 nm for BFO and ~20 nm for SRO thin films, respectively.

**Scanning-probe measurements and switching of ferroelectric polarizations of BFO thin films.** Scanning-probe measurements were carried out on a Digital Instruments Nanoscope-V Multimode AFM under ambient conditions. The surface topography, ferroelectric switching loops and domain structures were characterized using contact-mode AFM and PFM with commercially available Pt-coated Si tips (HQ:NSC35/PT, MikroMasch). During the measurements of topography and domain structures, typically scanning rate of the cantilever was 1 Hz and 0.2 Hz, respectively. The AC amplitude and frequency for ferroelectric domain and switching loop measurements were 0.5 V$_{pp}$ and 20 kHz, respectively. Pure downward and upward domains were written by applying $\pm 8$ V DC voltage on the scanning PFM tip with a grounded bottom electrode.

**The measurement of local I-V curves with/without optical excitation.** The local I-V curves were characterized with P-AFM, which composed with conductive atomic force microscopy (C-AFM) and a solar simulator with the power of 100 mw/cm$^2$. During the measurements, the contact force between the probe and film surface is ~30 nN [17], which would induce negligible strain and bandgap change [15] in the BFO thin film.

**Construction of capacitor structures.** Circular gold electrodes (radius, ~5 μm) were fabricated on top of the BFO thin films by e-beam lithography using an JEOL JBX-6300FS system. The positive Zep520A photoresist was spin coated on top of the BFO thin films. After exposure using the e-beam writer, the development was carried out in a ZED-N50 developer. Then the sample was rinsed with isopropyl alcohol and dried by nitrogen gas. At last, a 40-nm-thick gold layer was sputtered as the top electrodes using a lift-off process.

**The measurement of wavelength-dependent photocurrent in the large-area capacitor and point-contact geometry.** The wavelength dependent $I_{sc}$ was characterized by P-AFM in large-area capacitor and point-contact geometry with the intensity of ~15 mW/cm² in each wavelength. The general intensity of the optical absorption obtained from the point-contact geometry (red curve in Fig. 3(a)) is weaker than the counterpart from the capacitor structure (black curve in Fig. 3(a)) due to the much smaller accumulated area for photo-excited carriers.

**Phase-field simulations.** Ferroelectric polarization switching in BFO thin films was simulated using the phase-field model by solving the time-dependent Ginzburg-Landau (TDGL) equation for the temporal evolution of the polarization vector field,

$$\frac{\partial P_i(r,t)}{\partial t} = -L\frac{\delta F}{\delta P_i(r,t)}, i = 1,2,3 \qquad (1)$$

where $P_i(r, t)$ is the polarization, $L$ is a kinetic coefficient that is related to the domain wall mobility, and $F$ is the total free energy that includes contributions from the Landau energy, elastic energy, electric energy, and gradient energy.

$$F = \iiint (f_{Landau} + f_{elastic} + f_{electric} + f_{grad})dV \quad (2)$$

Detailed descriptions of the total free energy $F$ were presented in some depth in our previous publication [35]. The strain tensor trace ($\Sigma|s_{ii} - s'_{ii}|$) was calculated to estimate the inhomogeneous strain variation [36]. The Landau, elastic and electrostrictive coefficients used are described in the reference [37]. The equation was solved by a semi-implicit Fourier spectral method using a discreet grid of 256 Δx × 1 Δx × 128 Δx (Δx is the number of grid points and equals to 1 nm in this work) with the periodic boundary conditions along x and y directions in the film plane. A thin film was calculated by applying short-circuit boundary conditions in both film surfaces as described in detail in our previous publications [35, 38]. The thickness of 30 - 120 nm was used for the BFO thin films which was consistent with the experimental counterparts. When a photovoltage ($V_{oc}$) was applied, the distribution of the electric potential on the top surface was approximated by a Lorentz distribution [36],

$$\phi_L(x,y) = \phi_0(\frac{Y^2}{r^2+Y^2}) \qquad (3)$$

where $r$ is the lateral distance from the cone-shaped tip and $Y$ is the half-width at half-maximum.

**First-principle calculations.** We performed Density Functional Theory (DFT) calculations with the VASP package [39-40]. The exchange and correlation was generalized gradient approximation PBEsol+U (the optimized version of PBE [41]) and the DFT + U approach introduced by Liechtenstein [42] *et al*. We used $U$ = 7.0 eV and $J$ = 1.0 for the $d$ orbital of Fe. The projector-augmented wave (PAW) method [43] was used to describe the following electron states: Fe's 3$p$, 3$d$, and 4$s$; Bi's 5$d$, 6$s$, and 6$p$; and O's 2$s$ and 2$p$. Wave functions were represented in a plane-wave basis truncated at 500 eV. We adopt the 30-atom hexagonal unit cell and the 10-atom rhombohedral unit cell. In addition, G-type antiferromagnetic order was arranged to Fe spins. The electronic energies were convergent to at least 0.001 meV/atom, and all force components were relaxed to at least 2 meV/Å. The relaxations of cell shape and ions' positions were performed by the Gaussian smearing technique.

**Sample preparation and characterization using TEM.** The cross-section sample for *in-situ* TEM characterization was prepared by mechanically milling into a slice with a thickness of ~20 μm, transferring onto a half Cu grid and further thinning to electron transparent (~50 nm) with a focused ion beam (FIB). Pt was deposited on the SRO layer and connected with the Cu grid to make sure a good bottom electrode. The *in-situ* electric field experiments were performed on a Tecnai F20 TEM with a NanofactoryTM ST-1000 Holder and Gatan GIF system. A tungsten probe installed in this holder can touch BFO thin film and was used as a top electrode. EELS data were processed with the EELS package in Digital Micrograph.

**Force curves characterization.** Force curves were carried out to achieve an accurate contact force between the nanoscale probe and sample surface. Figure S8(a) in the supplementary information shows a force curve carried out on a hard substrate (silicon wafer), which is used to calibrate the deflection sensitivity ($D$, 35.96 nm/V). Taking advantage of the deflection sensitivity ($D$, 35.96 nm/V), the raw photodiode signal (in Volts) can be transformed to the deflection of the cantilever (in nm). Fig. S8(b) describes the force curve obtained on BFO thin film, where a setpoint value of -0.3 V was used. The contact force between the probe and the sample surface is defined by the equation:

$$F = k \times d \qquad (1),$$

where $k$ is the spring-constant of the cantilever and $d$ is the deflection measured from the setpoint to $V_{CSmin}$ (the point where the cantilever pulls off the surface) in nanometers. The $k$-value (~3.3 N/m) was derived from a thermal tune of the cantilever at multiple points. The deflection of the cantilever ($d$) was obtained by making use of the deflection sensitivity: $d$ = 0.64 V $\times$ 35.96 nm/V = 23 nm. Therefore, the contact force was calculated as: $F$ = 3.3 N/m $\times$ 23 nm = 75 nN.

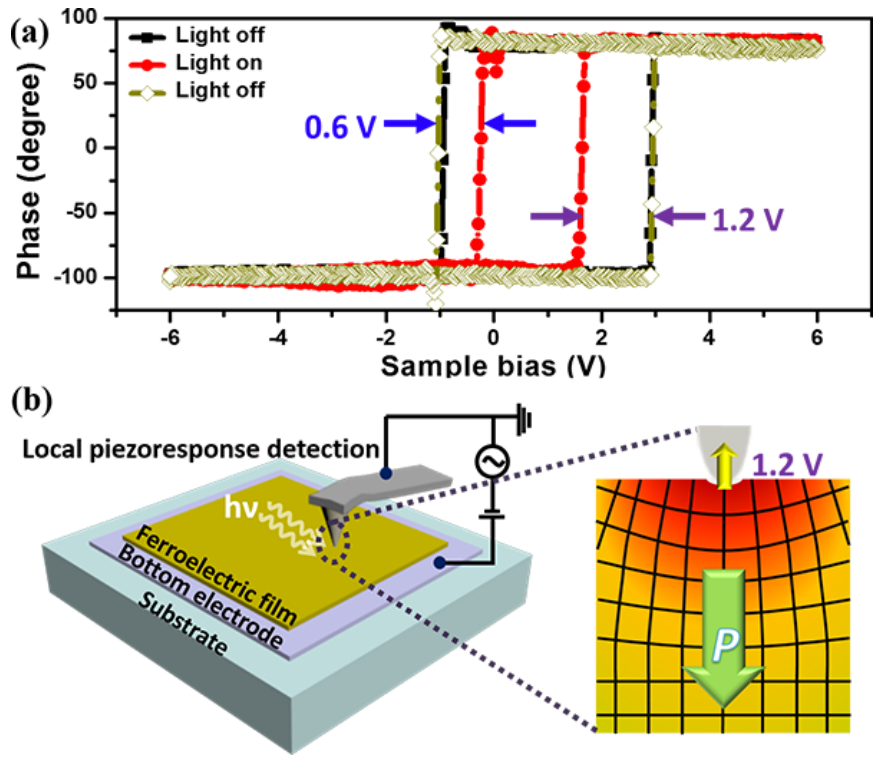

**Figure 1.** Visible light-excited probe voltage and the local strain at the probe/film interface. (a) Local ferroelectric switching loops without (black and brown curves)/with (red curve) the presence of visible light at the probe/film interface. (b) The schematic of experimental setup for local piezoresponse measurement (left panel) and the visible light-excited probe voltage (yellow arrow) induced local strain in ferroelectric thin film with downward polarizations (green arrow).

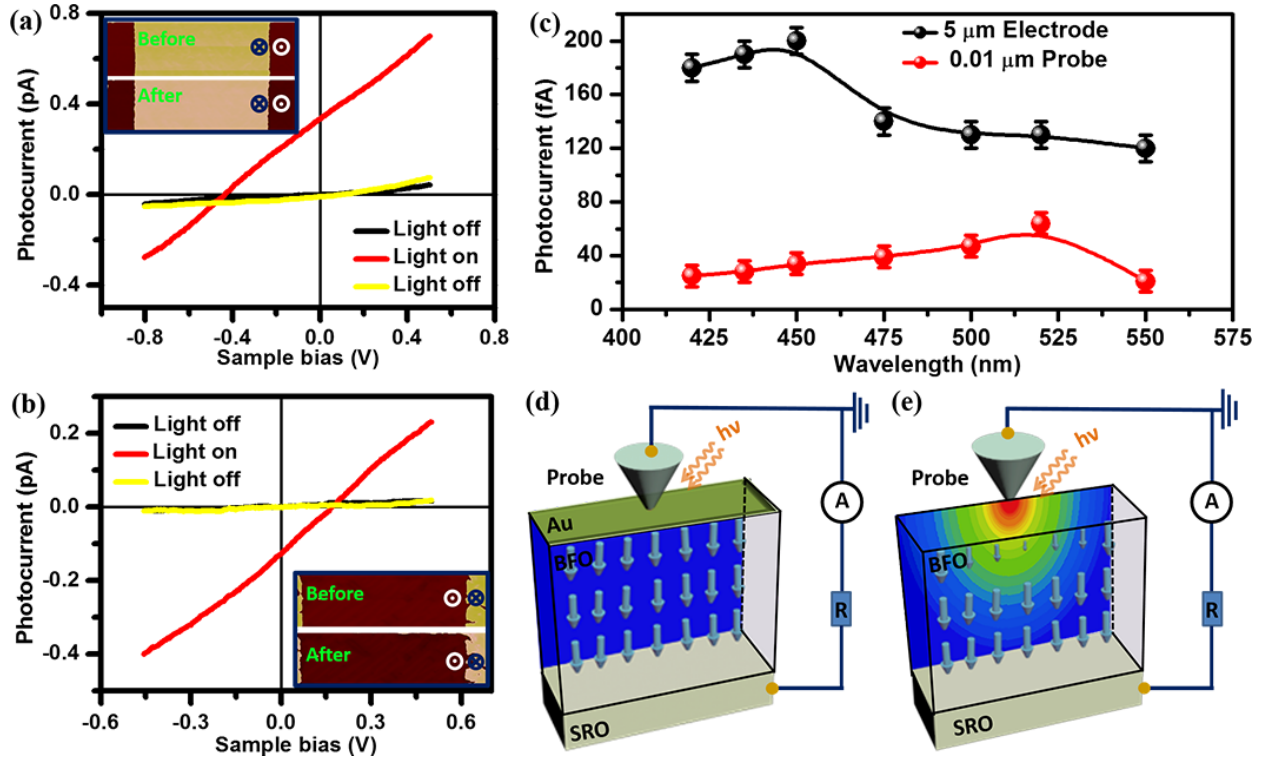

**Figure 2.** Switchable photovoltaic effects and local bandgap variation in BFO thin film based on a point-contact geometry. Transparent-conductive tip (Dcp 20, NT-MDT) was employed to achieve efficient optical absorption and local transport measurements across the interface. (a), (b) Local I-V curves in the dark (black and yellow curves) and under illumination (red curves, 100 mW/cm$^2$) obtained from the single domain region with downward (a) and upward (b) polarizations. The inserted out-of-plane PFM images in (a) and (b) demonstrate that the polarizations did not change after the local transport measurements. The scanning size of out-of-plane PFM images is 5.8×1.5 μm$^2$. (c) Spectral responses of $I_{sc}$ obtained from a capacitor structure (black curve) and point-contact geometry (red curve), respectively. (d), (e) Schematics of the mechanism of the nanoscale bandgap drop in BFO thin film based on the point-contact geometry (d) compared with the capacitor structure (e) respectively.

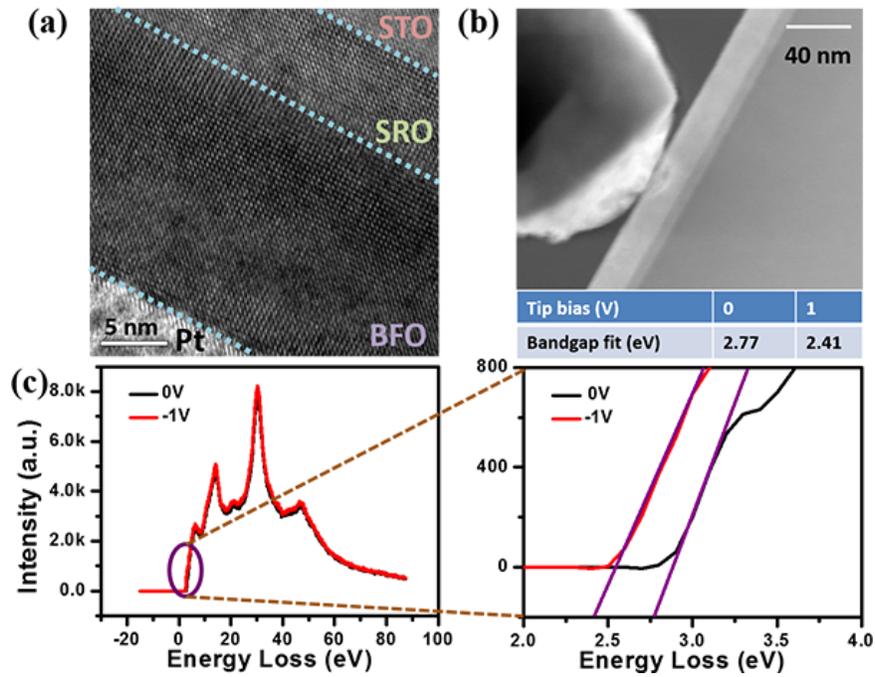

**Figure 3.** Nanoscale bandgap drop in BFO thin film with the application of probe voltage. (a) High resolution transition electron microscopy (TEM) image for the epitaxial hetero-structure of BFO/SRO/STO(001). (b) Point-contact geometry based on TEM. (c) *In-situ* EELSs obtained in BFO thin film with out-of-plane downward polarizations at the location close to the contact point without (black curve)/with (red curve) the application of 1V probe voltage (upward) respectively. The linear fitting of the onset region of EELSs in the magnified image shows a bandgap drop (~0.36 eV) in the biased state compared to the initial state.

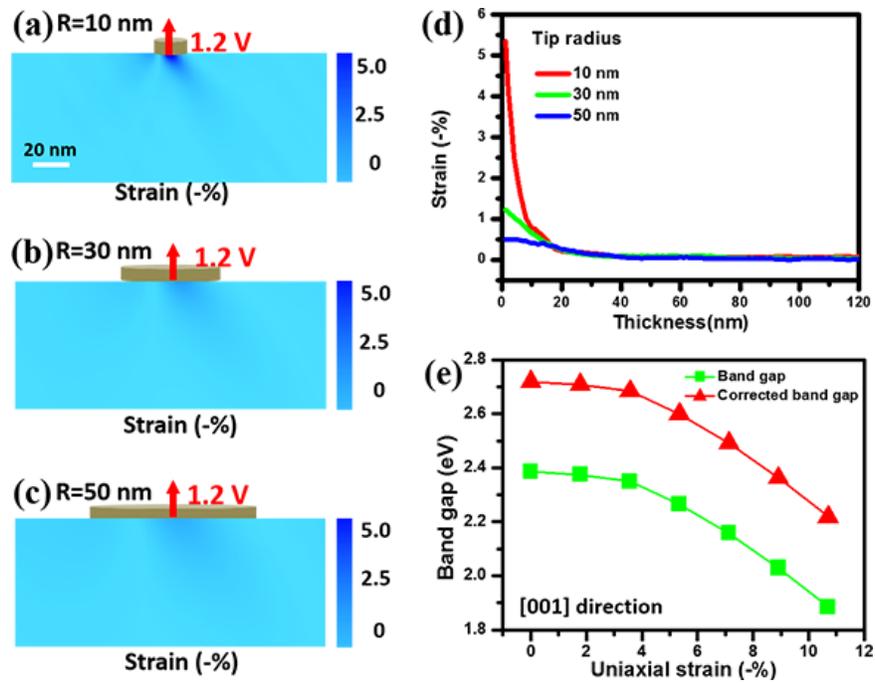

**Figure 4.** Quantitative analysis of the strain and nanoscale bandgap drop in BFO thin film (~120 nm). Phase-field simulations of the strain in BFO thin film with out-of-plane downward polarizations in point-contact geometries with the application of 1.2 V (upward) probe voltage with 10 nm (a), 30 nm (b) and 50 nm (c) contact radius, respectively. (d) Strain along the thickness direction of the BFO thin film underneath the probe with various contact radius. (e) First-principle calculations for the bandgap of BFO thin film under an increasing compressive strain along [001] direction.

ASSOCIATED CONTENT

**Supporting Information**.

The Supporting Information is available free of charge on the ACS Publications website.

AFM, XRD, SEM, PFM, local photo-electronic measurement, phase-field simulations, first-principle calculations and force curve measurement.

AUTHOR INFORMATION

**Corresponding Author**

*E-mail: jxzhang@bnu.edu.cn

**Author Contributions**

[#] These authors contributed equally to this work. J. W. performed the experiments. H. H. performed the phase-field simulations. W. H. performed DFT calculations. Q. Z. performed *in-situ* EELSs. D. Y., Y. Z. R. L. and C. W. performed the fabrication and characterization of materials. All the authors discussed and commented on this study. J. W., H. H. and J. Z. designed the experiments and wrote the manuscript.

**Notes**

The authors declare no competing financial interest.

**ACKNOWLEDGEMENTS**

The work at Beijing Normal University is supported by the NSFC under Contract 51332001 and the National Key Research and Development Program of China through Contract 2016YFA0302300. J.Z. also acknowledges the support from the National Basic Research Program of China, under Contract 2014CB920902. H. H. acknowledges the support from NSFC under Contract 11504020. The work at Beijing Normal University is also supported by "The Fundamental Research Funds for the Central Universities" under Contracts 2017EYT26 and 2017STUD25. The effort at Penn State is supported by the U.S. Department of Energy, Office of Basic Energy Sciences, Division of Materials Sciences and Engineering under Award FG02-07ER46417.

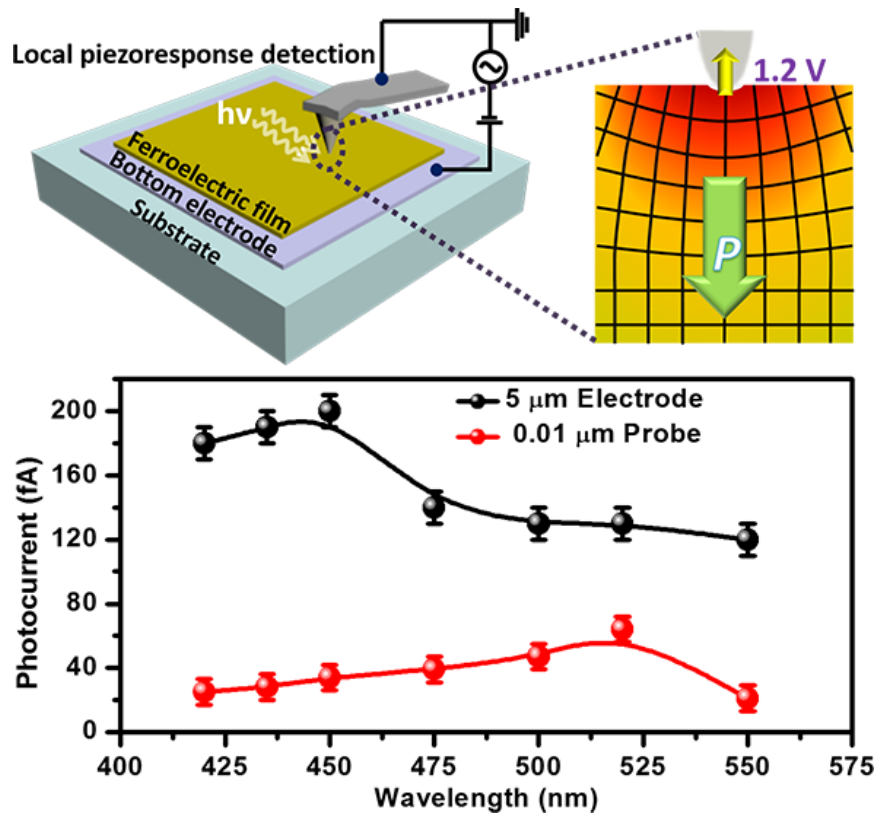